\documentclass[twocolumn]{webofc}

\usepackage[varg]{txfonts}   
\usepackage{graphicx}
\usepackage{caption} 
\usepackage{subcaption}
\usepackage{wrapfig} 
\usepackage{hyperref}
\usepackage{url}

\usepackage{titlesec}

\titlespacing*{\section}
{0pt}{0.8ex plus 0.2ex minus 0.2ex}{0.5ex plus 0.2ex minus 0.2ex}

\titlespacing*{\subsection}
{0pt}{0.6ex plus 0.2ex minus 0.2ex}{0.4ex plus 0.2ex minus 0.2ex}

\usepackage{etoolbox}
\patchcmd{\thebibliography}{\small}{\footnotesize}{}{}
\hypersetup{colorlinks=true,citecolor=blue,urlcolor=blue,linkcolor=blue}
\begin{document}
\title{Ion Optics for quasi-free $(p,p\alpha)$ reactions with Grand Raiden Spectrometer}
%
%

\author{\firstname{Taichi} \lastname{Miyagawa}\inst{1,2}\fnsep\thanks{\email{miyatai@rcnp.osaka-u.ac.jp}} 
\and
        \firstname{Junki} \lastname{Tanaka}\inst{1,2}\fnsep
        \firstname{for} \lastname{the ONOKORO collabration}
}

\institute{Research Center for Nuclear Physics (RCNP), The University of Osaka, Osaka 567-0047, Japan
\and
Department of Physics, The University of Osaka, Osaka 560-0043, Japan}

\abstract{
The quasi-free $(p,p\alpha)$ reaction is a powerful tool to probe preformed $\alpha$ clusters in nuclei, but it requires accurate reconstruction of both momentum and scattering angles at the reaction point. In this work, ion-optical analysis for $(p,p\alpha)$ measurements with the Grand Raiden spectrometer is presented. An under-focus optical setting was adopted to preserve sensitivity to the vertical scattering angle while maintaining high momentum resolution.
The focal-plane geometry was determined independently as a purely geometrical reference.
Momentum calibration was performed using elastic scattering of $^{206}$Pb$(p,p)$ at a fixed spectrometer angle.
Scattering angles were reconstructed using ion-optical relations, and residual higher-order effects were corrected by a multidimensional fit.
The dominant contributions to the reconstruction were found to arise from terms up to third order within the experimental acceptance.
This ion-optical framework enables consistent event-by-event reconstruction of the reaction kinematics and provides a reliable basis for quasi-free $(p,p\alpha)$ analyses.
}

\maketitle

\section{Ion-Optics for $(p,p\alpha)$ Measurements}

The formation of $\alpha$ clusters in nuclei is a key issue in nuclear structure physics, particularly in the low-density surface region where clustering is expected to emerge.
Quasi-free $(p,p\alpha)$ reactions provide a direct probe of such cluster structures by selectively knocking out preformed $\alpha$ clusters.
In order to isolate quasi-free processes and to reconstruct the internal momentum of the knocked-out cluster, an accurate reconstruction of the reaction kinematics at the interaction point is essential.

In $(p,p\alpha)$ measurements, the reaction kinematics are determined from the momentum and scattering angles of the detected proton and $\alpha$ particle.
For the proton arm, a high-resolution magnetic spectrometer is typically employed to achieve the required momentum resolution. However, ion-optical settings optimized for momentum spectroscopy do not necessarily preserve sufficient sensitivity to all angular degrees of freedom at the focal plane. As a result, a configuration that is adequate for inclusive momentum measurements may not be suitable for exclusive $(p,p\alpha)$ reactions.

A particularly important limitation arises when the vertical direction of the spectrometer is tuned to be double-focused at the focal plane. While this condition maximizes momentum resolution, it suppresses the dependence of focal-plane observables on the vertical scattering angle at the target.
For $(p,p\alpha)$ reactions, where the full reconstruction of the four-momentum transfer is required to identify quasi-free kinematics, this loss of angular information becomes a serious drawback.

To address this issue, the present work adopts an under-focus ion-optical configuration of the Grand Raiden spectrometer.
By intentionally relaxing the vertical focusing condition, the focal-plane observables retain measurable sensitivity to the scattering angles while maintaining high momentum resolution. This optical setting allows momentum and scattering angles to be reconstructed simultaneously and consistently, which is a prerequisite for reliable $(p,p\alpha)$ analysis.
The ion-optical analysis presented in this work is therefore organized according to the logical structure of the reconstruction problem. First, the focal-plane geometry is determined as a purely geometrical reference, independent of reaction kinematics.
Second, the momentum calibration is performed using elastic scattering under well-defined conditions.
Finally, the scattering angles are reconstructed using ion-optical relations, including higher-order terms determined by a multidimensional fit.
This separation ensures that each reconstruction step is well defined and that systematic correlations between different stages are minimized.

\section{Focal-plane Reconstruction}

\begin{figure}[h!]
    \centering
    \includegraphics[width=\linewidth]{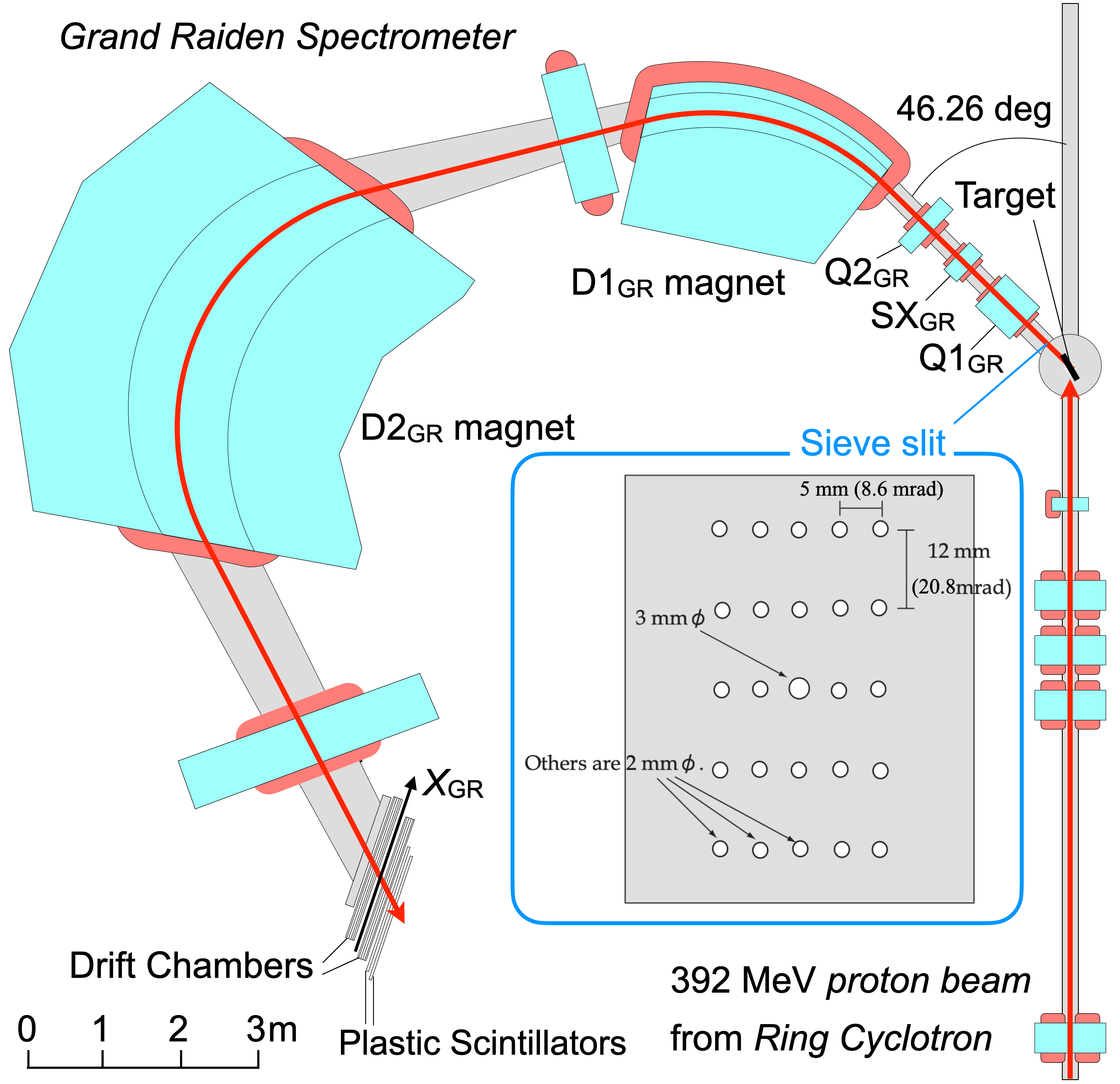}
    \caption{Schematic view of the Grand Raiden spectrometer. The set angle 28 degrees is for the sieve slit calibration measurements. Insert shows the sieve slits used for angle reconstruction.}
    \label{fig:setup_sieve}
\end{figure}

The ion-optical analysis was performed using the Grand Raiden (GR) high-resolution
magnetic spectrometer at RCNP.
GR consists of a Q1--Q2--SX--D1--D2 magnet system and provides a momentum resolution
of $\Delta p/p = 1/37000$.
A schematic view of the spectrometer and detector system is shown in
Fig.~\ref{fig:setup_sieve}.

The focal plane is defined as the geometrical plane where the horizontal position
exhibits the maximum intrinsic resolution with minimal correlation to the
reaction-point coordinates and scattering angles.
This definition is purely geometrical and independent of the momentum and angular
reconstructions.
Downstream of the GR exit, particle trajectories are approximated as linear.
Accordingly, the horizontal positions $x_1$ and $x_2$ measured at two parallel
planes separated by a distance $z$ satisfy
\begin{equation}
x_1 = x_2 + a_2 z ,
\label{eq:fp_linear}
\end{equation}
where $a_2$ is the local trajectory angle in the dispersive plane.
Equation~\eqref{eq:fp_linear} allows the detector-plane distributions to be
projected onto arbitrary virtual planes.

\begin{figure}[h!]
    \centering
    \includegraphics[width=0.5\textwidth]{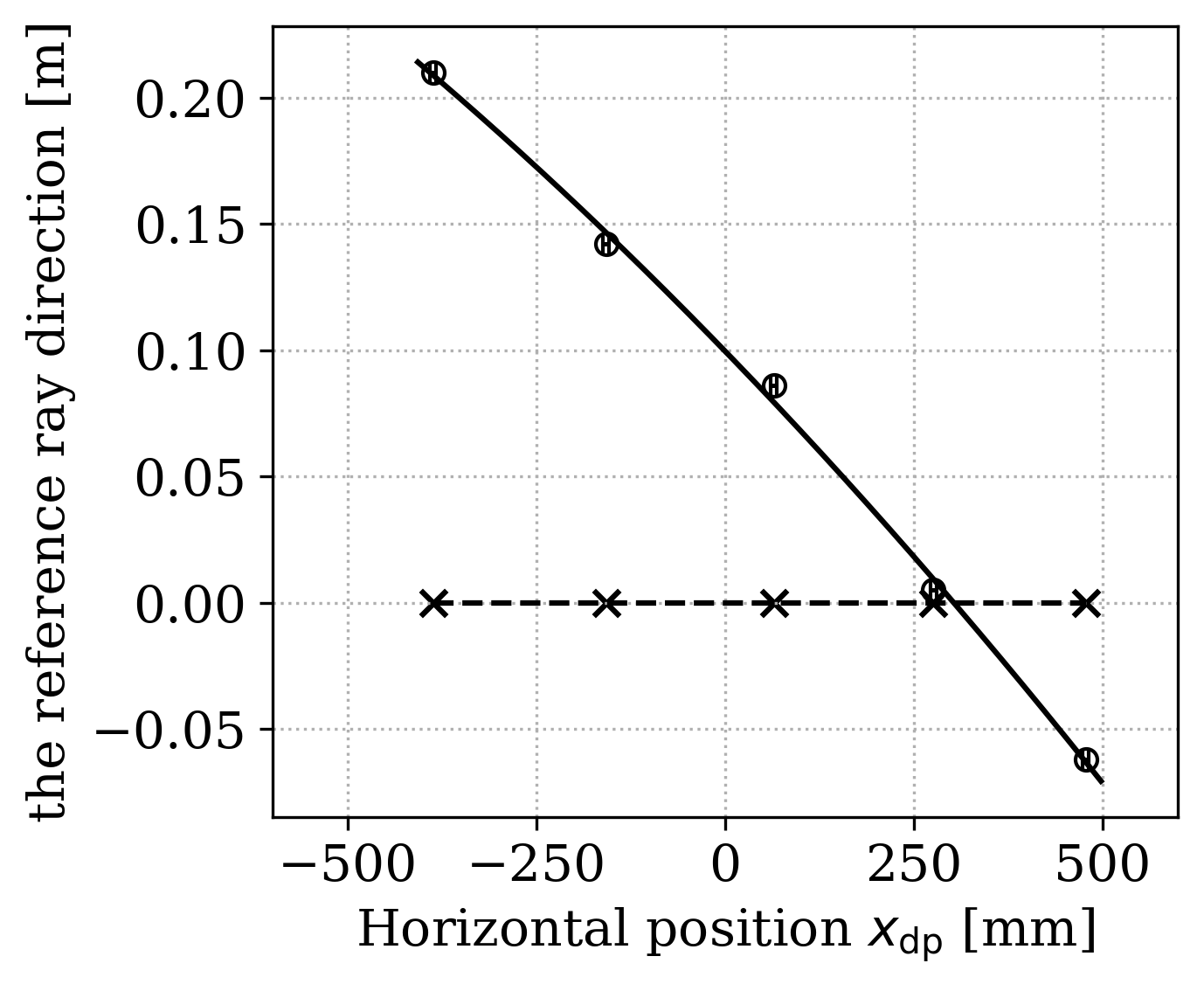}
    \caption{Determination of the GR focal plane. The dashed line indicates the detector plane, and the solid line the focal plane defined by the minimum horizontal width.}
    \label{Fig:GRFP_z}
\end{figure}

By varying $z$ and minimizing the width of the projected horizontal peak
distributions, the focal-plane position was determined experimentally without
relying on ion-optical model assumptions.
The result is shown in Fig.~\ref{Fig:GRFP_z}.
The distance between the focal plane and the detector system is parameterized as a
function of the horizontal position $x$ by
\begin{equation}
z = -6.70\times10^{-7}x^2 - 3.0829\times10^{-4}x + 0.0995\ [\mathrm{m}],
\label{eq:fp_prm}
\end{equation}
which constitutes an independent parameter set not shared with the momentum or
angular reconstruction procedures.

Once fixed, the focal-plane position was kept unchanged throughout the analysis,
and all focal-plane observables were defined at this plane by projection from the
detector plane using Eq.~\eqref{eq:fp_linear}.

\section{Momentum Reconstruction}

After fixing the focal-plane geometry, the momentum of the scattered proton is
reconstructed from the horizontal focal-plane position $x_{fp}$ using an
independent set of dispersive ion-optical parameters.
The momentum reconstruction is therefore independent of the focal-plane
determination.

The focal-plane coordinate $x_{fp}$ is expanded around the central trajectory as
\begin{equation}
\begin{aligned}
x_{fp}
={}& x_{fp0}
+ \left( \frac{\partial x_{fp}}{\partial x_{sc}} \right)_c x_{sc}
+ \left( \frac{\partial x_{fp}}{\partial a_{sc}} \right)_c a_{sc} \\
&+ \left( \frac{\partial x_{fp}}{\partial y_{sc}} \right)_c y_{sc}
+ \left( \frac{\partial x_{fp}}{\partial b_{sc}} \right)_c b_{sc}
+ \left( \frac{\partial x_{fp}}{\partial \delta} \right)_c \delta .
\end{aligned}
\end{equation}

In the present experiment, the incident beam was transported achromatically to the
target, resulting in $x_{tgt} \simeq 0$ and $y_{tgt} \simeq 0$, and the symmetry of
the magnetic field suppresses the dependence on the scattering angles.
Accordingly, the focal-plane position is dominated by the momentum deviation
$\delta$ and is well approximated by
\begin{equation}
x_{fp}
=
x_{fp0}
+ \left( \frac{\partial x_{fp}}{\partial \delta} \right)_c \delta .
\label{eq:xfp_delta}
\end{equation}

The momentum calibration was performed using elastic scattering protons from the
$^{206}$Pb$(p,p)$ reaction with the GR spectrometer fixed at a laboratory
scattering angle of 28~deg.
By varying the magnetic rigidity, the same process was measured at different
momentum deviations $\delta$.

\begin{figure}[h!]
    \centering
    \includegraphics[width=0.5\textwidth]{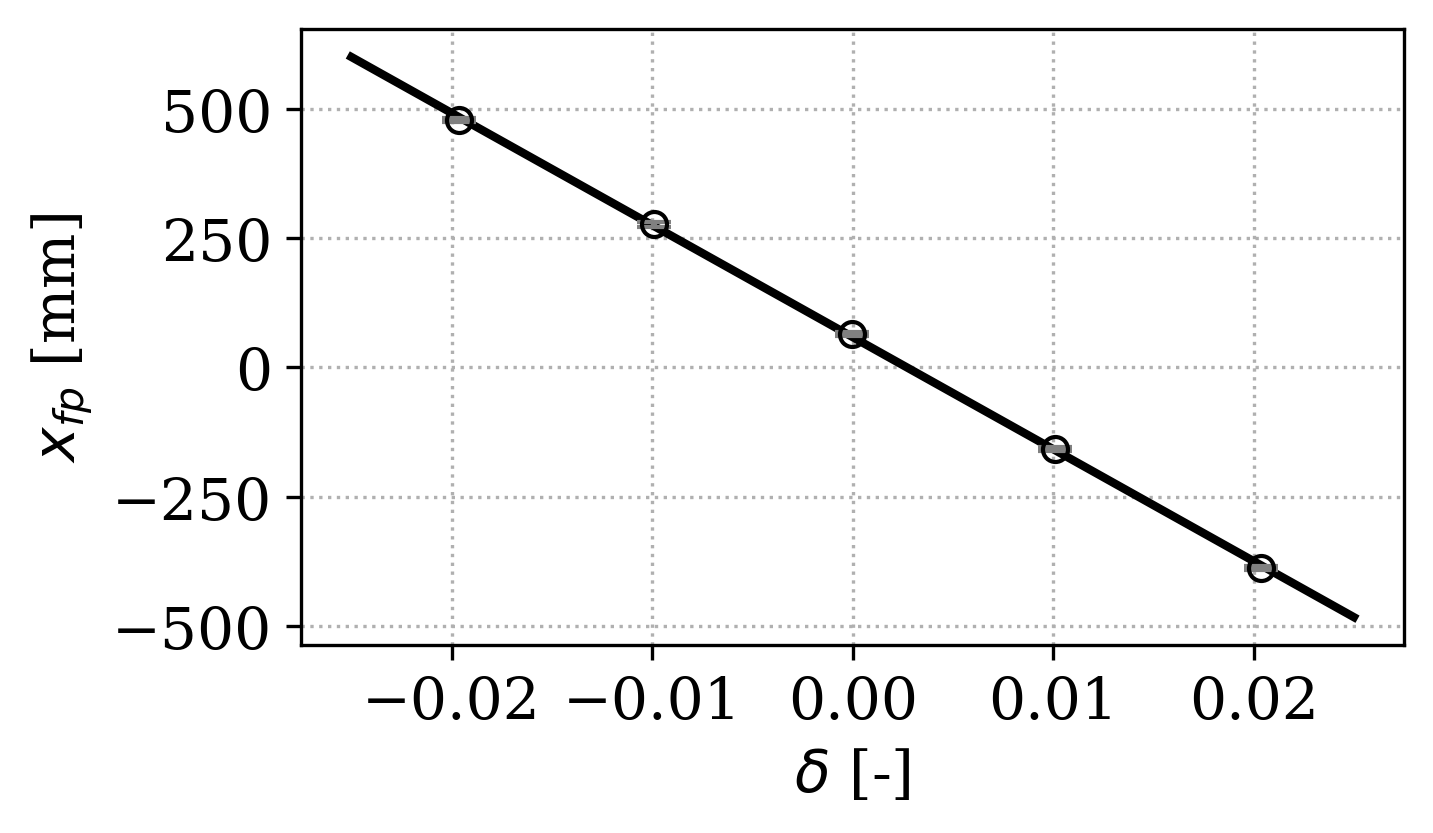}
    \caption{Correlation between the focal-plane position $x_{fp}$ and the momentum
    deviation $\delta$ obtained from the $^{206}$Pb$(p,p)$ calibration.}
    \label{Fig:GRFP_xfp_delta}
\end{figure}

Figure~\ref{Fig:GRFP_xfp_delta} shows the correlation between $x_{fp}$ and
$\delta$, from which the dispersive coefficient
$\left( \partial x_{fp} / \partial \delta \right)_c$ and the reference position
$x_{fp0}$ were determined by a linear fit.

The momentum deviation is reconstructed event by event using the inverse relation
of Eq.~\eqref{eq:xfp_delta},
\begin{equation}
\delta
=
\frac{x_{fp}-x_{fp0}}
{\left( \partial x_{fp}/\partial\delta \right)_c}.
\label{eq:delta_recon}
\end{equation}

The parameters required for momentum reconstruction are summarized in
Table~\ref{tab:momentum_param} and kept fixed throughout the subsequent angular
reconstruction.

\begin{table}[h!]
\centering
\caption{Parameters for momentum reconstruction used in the present analysis.}
\label{tab:momentum_param}
\begin{tabular}{lc}
\hline\hline
Parameter & Value \\
\hline
Reference focal-plane position $x_{fp0}$ & $58.95$ mm \\
Dispersive coefficient $\left( \partial x_{fp} / \partial \delta \right)_c$
& $-21583.23$ mm \\
\hline\hline
\end{tabular}
\end{table}

\section{Angular Reconstruction}

The reconstruction of the scattering angles was performed using a right-handed
coordinate system defined along the central trajectory of the GR spectrometer.
The horizontal and vertical scattering angles at the target were defined as
$A = dZ/dX$ and $B = dZ/dY$, respectively, and the momentum deviation as
$\delta = (p - p_0)/p_0$.
The beam spot at the target position was approximately 1~mm in FWHM.

The angular reconstruction was carried out after fixing the focal-plane geometry
and the momentum reconstruction parameters.
In this step, the focal-plane observables are related to the scattering angles at
the target through the ion-optical response of the spectrometer.

The first-order reconstruction formulas are given by
\begin{equation}
A_{\mathrm{sc}} =
\frac{A_{\mathrm{fp}} - (a|\delta)\,\delta}{(a|a)},
\label{eq:Asc}
\end{equation}
\begin{equation}
B_{\mathrm{sc}} =
\frac{Y_{\mathrm{fp}}}{
(y|b) + (y|b\delta)\,\delta + (y|ab)\,A_{\mathrm{sc}} },
\label{eq:Bsc}
\end{equation}
where the ion-optical coefficients describe the response of the focal-plane
observables to the scattering angles and the momentum deviation.

\begin{figure}[h!]
    \centering
    \includegraphics[width=\linewidth]{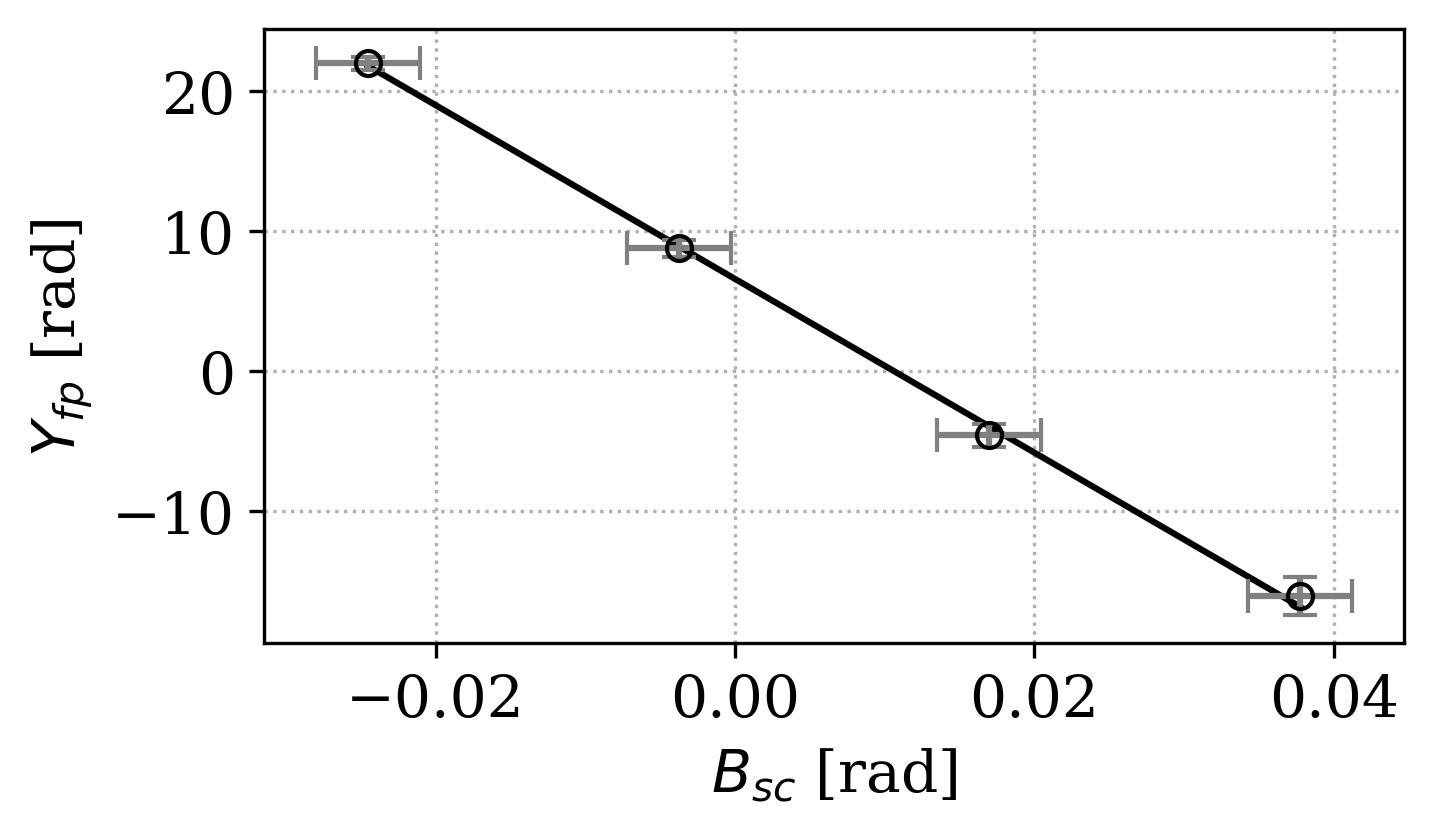}
    \caption{Ion-optical correlation $Y_{\mathrm{fp}}$ vs.\ $B_{\mathrm{sc}}$
    used to determine $(y|b)$.}
    \label{fig:aa_yb}
\end{figure}

The coefficients entering Eqs.~\eqref{eq:Asc} and~\eqref{eq:Bsc} were extracted
sequentially from one-dimensional correlations.
From the $A_{\mathrm{sc}}$--$A_{\mathrm{fp}}$ correlation
(Fig.~\ref{fig:aa_yb}(a)), the horizontal-angle magnification $(a|a)$ was obtained
as the slope, while the offset provides a constraint on the dispersive term
$(a|\delta)$.
The vertical-angle response coefficient $(y|b)$ was determined from the
$B_{\mathrm{sc}}$--$Y_{\mathrm{fp}}$ correlation
(Fig.~\ref{fig:aa_yb}(b)).
Because the under-focus optics restore the sensitivity of $Y_{\mathrm{fp}}$ to
$B_{\mathrm{sc}}$, the slope of this correlation is a key parameter for
reconstructing the vertical scattering angle.

\begin{figure}[h!]
    \centering
    \includegraphics[width=\linewidth]{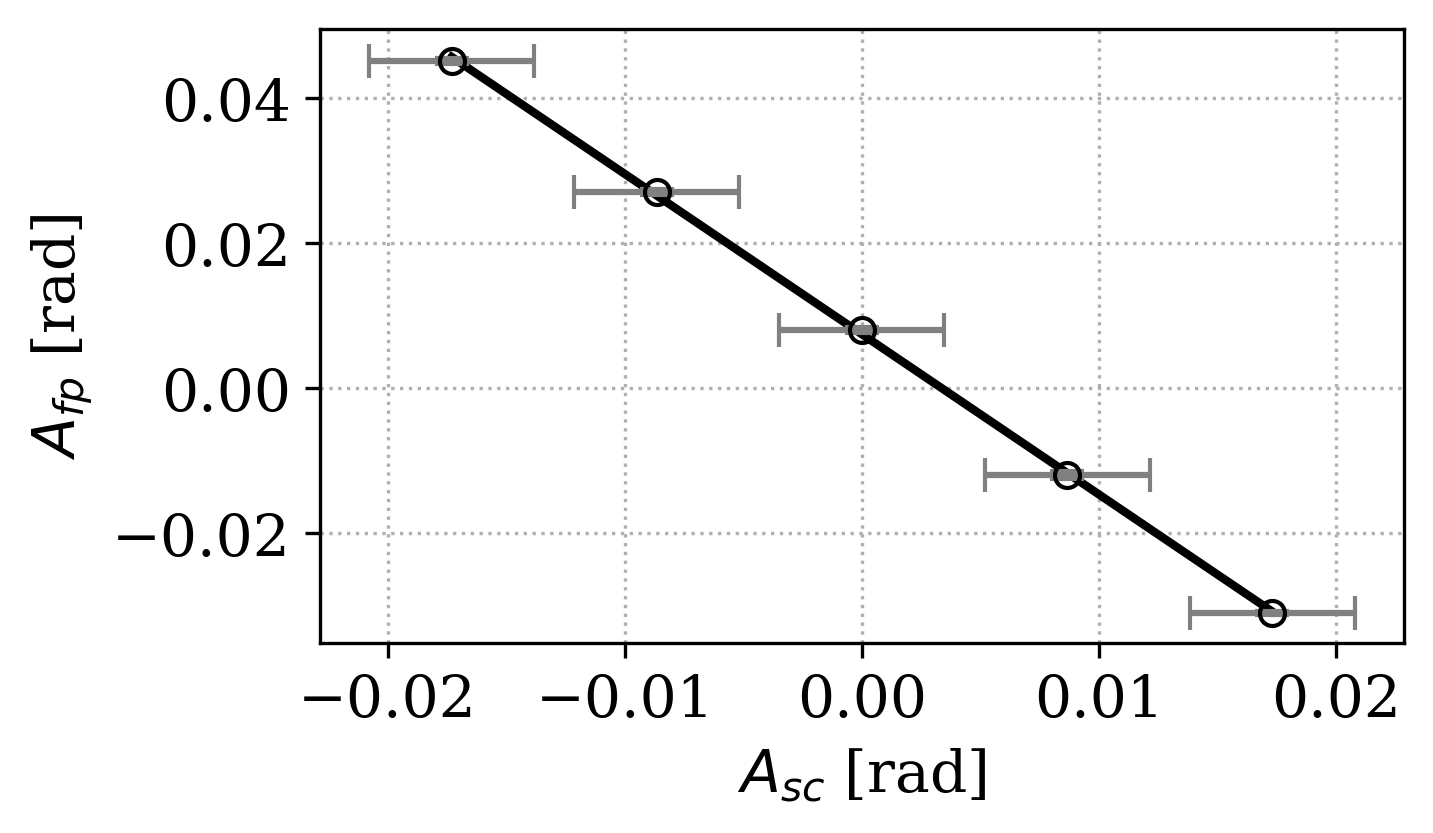}
    \caption{Ion-optical correlation $A_{\mathrm{fp}}$ vs.\ $A_{\mathrm{sc}}$
    used to determine $(a|a)$.}
\end{figure}

\begin{figure}[h!]
    \centering
    \includegraphics[width=\linewidth]{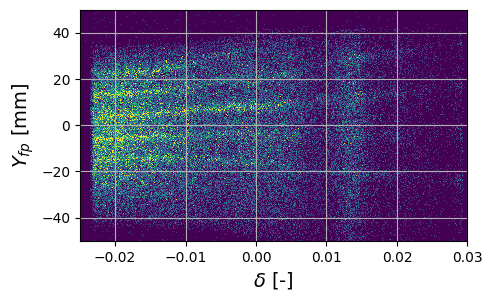}
    \caption{$Y_{\mathrm{fp}}$ vs.\ $\delta$ correlation for $(y|b\delta)$.}
    \label{fig:ybd}
\end{figure}

Higher-order distortions in the GR ion optics modify the apparent slopes of the
first-order coefficients.
To quantify the momentum dependence of the angular response, correlations between
the focal-plane observables and the momentum deviation were examined.
Figure~\ref{fig:ybd} shows the correlations used to extract $(a|\delta)$ and
$(y|b\delta)$.

\begin{figure}[h!]
    \centering
    \includegraphics[width=\linewidth]{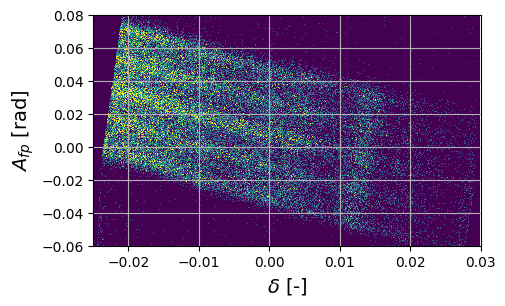}
    \caption{$A_{\mathrm{fp}}$ vs.\ $\delta$ correlation for $(a|\delta)$.}
\end{figure}

Figures~\ref{fig:aa_yb} and~\ref{fig:ybd} provide the experimentally determined
constraints on the dominant first-order and leading momentum-dependent
coefficients.
The experimentally determined results from the one-dimensional and dominant
two-dimensional correlations are summarized in Table~\ref{tab:angular_coeff}.

\begin{table}[h!]
\centering
\caption{Dominant first-order and momentum-dependent ion-optical coefficients}
\label{tab:angular_coeff}
\begin{tabular}{lc}
\hline\hline
Coefficient & Value \\
\hline
$(a|a)$        & $-2.21$ [-] \\
$(a|\delta)$   & $-1.199$ [rad] \\
$(y|b)$        & $-614.319$ [mm/rad] \\
$(y|b\delta)$  & $-8443.49$ [mm/rad] \\
$(y|ab)$       & $4637.437$ [mm/$\mathrm{rad}^2$] \\
\hline\hline
\end{tabular}
\end{table}

\begin{figure}[h!]
    \centering
    \includegraphics[width=\linewidth]{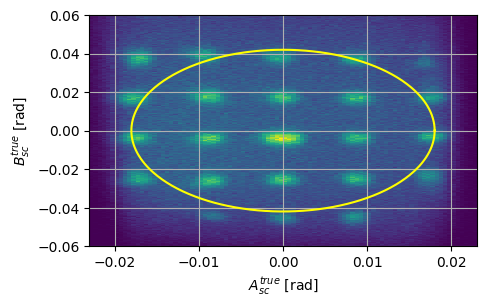}
    \caption{Reconstructed sieve pattern after multidimensional inverse mapping.
    The yellow ellipse denotes the solid-angle acceptance used in this analysis.}
    \label{fig:recon_compare}
\end{figure}

Figure~\ref{fig:recon_compare}(a) shows the sieve pattern reconstructed using the
coefficients listed in Table~\ref{tab:angular_coeff}.
Although the overall hole pattern is reproduced, residual deviations from the
geometrical layout remain, suggesting the presence of uncorrected higher-order
terms.
Based on these results, the final inverse ion-optical mapping was obtained through
a multidimensional fit to the sieve-slit data.

In this procedure, a model function including linear and mixed terms was
constructed using the first-order scattering angles $A_{\mathrm{sc}}$ and
$B_{\mathrm{sc}}$ derived from Eqs.~\eqref{eq:Asc} and~\eqref{eq:Bsc},
together with the momentum deviation $\delta$.
The model coefficients were optimized to best reproduce the true scattering
angles ($A_{\mathrm{sc}}^{\mathrm{true}}, B_{\mathrm{sc}}^{\mathrm{true}}$)
determined from the sieve-slit geometry.
As a result, the coupled dependence of $A_{\mathrm{fp}}$, $Y_{\mathrm{fp}}$, and
$\delta$ over the full acceptance was successfully reproduced.

Although the inverse mapping is formulated in a general polynomial form, the
dominant contributions within the experimental acceptance were found to arise
mainly from terms up to third order.
As shown in Fig.~\ref{fig:recon_compare}(b), the reconstructed hole positions
coincide with their geometrical locations, demonstrating that the remaining
distortions are effectively removed.

\begin{figure}[h!]
    \centering
    \includegraphics[width=\linewidth]{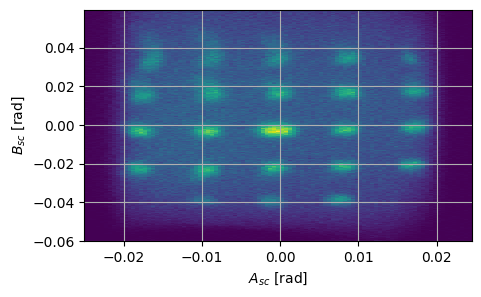}
    \caption{Reconstructed sieve pattern using analytical optical coefficients.}
\end{figure}

The angular resolution was evaluated from the reconstructed
$A_{\mathrm{sc}}^{\mathrm{true}}$ and $B_{\mathrm{sc}}^{\mathrm{true}}$
distributions of the sieve--slit events.
The resulting $1\sigma$ resolutions are
$1.38 \pm 0.05~\mathrm{mrad}$ in the horizontal direction and
$2.60 \pm 0.09~\mathrm{mrad}$ in the vertical direction.
The reconstruction accuracy, quantified by the differences between the true and
reconstructed slit-hole angles, yields standard deviations of
$\Delta A = 0.26~\mathrm{mrad}$ and $\Delta B = 1.2~\mathrm{mrad}$.
These uncertainties were propagated to the semi-minor and semi-major axes of the
elliptical acceptance shown in Fig.~\ref{fig:recon_compare}(b), centered at
$(A_{\mathrm{sc}}^{\mathrm{true}}, B_{\mathrm{sc}}^{\mathrm{true}}) = (0,0)$ with
nominal axes of 18~mrad and 42~mrad, respectively.
The corresponding solid angle is $2.38 \pm 0.05~\mathrm{sr}$.

\section{Conclusion}

An ion-optical analysis for quasi-free $(p,p\alpha)$ measurements was established
using the GR spectrometer. An under-focus optical configuration was adopted to retain sensitivity to
scattering angles while preserving high momentum resolution.
The analysis consisted of three steps.
First, the focal-plane geometry was determined as a geometrical reference.
Second, the momentum reconstruction was calibrated using elastic
$^{206}$Pb$(p,p)$ scattering at a fixed angle.
Finally, the scattering angles were reconstructed using ion-optical relations,
with higher-order aberrations corrected by a multidimensional inverse mapping.
This separation minimizes systematic correlations and allows independent
validation.
The resulting ion-optical coefficients up to third order enable event-by-event
kinematic reconstruction over the full acceptance.
This framework provides a robust basis for identifying quasi-free $(p,p\alpha)$
events and extracting missing-momentum distributions with controlled systematic
uncertainties for studies of $\alpha$-cluster formation in medium-mass nuclei.



\vspace{2pt}
\noindent\textbf{Acknowledgments} \;
The authors thank the accelerator group at RCNP for their technical support. 
This work was supported by JSPS KAKENHI (JP21H04975) and the JSPS A3 Foresight Program ``Nuclear Physics in the 21st Century.'' 
Additional support was provided by the NRF TOPTIER Korea--Japan Joint Research Program (RS-2024-00436392), IBS (IBS-R031-D1) in Korea, and the National Key R\&D Program of China (2023YFE0101500).
\vspace{-10pt}
\end{document}